**Article Type: Hot Topic**

# Adapting to the Impact of AI in Scientific Writing: Balancing Benefits and Drawbacks while Developing Policies and Regulations

**Ahmed S. BaHammam, MD**[1,2]

[1]Department of Medicine, University Sleep Disorders Center, and Pulmonary Service, King Saud University, Riyadh, KSA, Saudi Arabia

[2]The Strategic Technologies Program of the National Plan for Sciences and Technology and Innovation in the Kingdom of Saudi Arabia, Riyadh, Saudi Arabia

ashammam2@gmail.com

https://orcid.org/0000-0002-1706-6167

**Khaled Trabelsi**[3]

[3]High Institute of Sport and Physical Education of Sfax, University of Sfax, Sfax 3000, Tunisia
Research Laboratory: Education, Motricity, Sport and Health, EM2S, LR19JS01, University of Sfax, Sfax 3000, Tunisia

trabelsikhaled@gmail.com

ORCID: https://orcid.org/0000-0003-2623-9557

**Seithikurippu R. Pandi-Perumal**[4,5]

[4]Saveetha Medical College and Hospitals, Saveetha Institute of Medical and Technical Sciences, Saveetha University, Chennai 602105, Tamil Nadu, India

[5]Division of Research and Development, Lovely Professional University, Phagwara, Punjab-144411, India
Pandiperumal2023@gmail.com

ORCID: https://orcid.org/0000-0002-8686-7259

**Haitham Jahrami**

[6]Government Hospitals, Manama, Bahrain

[7]Department of Psychiatry, College of Medicine and Medical Sciences, Arabian Gulf University, Manama, Bahrain
haitham.jahrami@outlook.com

ORCID: https://orcid.org/0000-0001-8990-1320

**Running title:** Benefits and Drawbacks of AI in Scientific Writing

**Corresponding author's full contact details:**
**Prof. Ahmed BaHammam**
Professor of Medicine
University Sleep Disorders Center, Department of Medicine, College of Medicine, King Saud University
Box 225503, Riyadh 11324, Saudi Arabia
Telephone: 966-11-467-9495
Fax: 966-11-467-9179
E-mails: ashammam2@gmail.com

**Abstract:**

This article examines the advantages and disadvantages of Large Language Models (LLMs) and Artificial Intelligence (AI) in research and education and proposes the urgent need for an international statement to guide their responsible use. LLMs and AI demonstrate remarkable natural language processing, data analysis, and decision-making capabilities, offering potential benefits such as improved efficiency and transformative solutions. However, concerns regarding ethical considerations, bias, fake publications, and malicious use also arise. The objectives of this paper are to critically evaluate the utility of LLMs and AI in research and education, call for discussions between stakeholders, and discuss the need for an international statement. We identify advantages such as data processing, task automation, and personalized experiences, alongside disadvantages like bias reinforcement, interpretability challenges, inaccurate reporting, and plagiarism. Stakeholders from academia, industry, government, and civil society must engage in open discussions to address the ethical, legal, and societal implications. The proposed international statement should emphasize transparency, accountability, ongoing research, and risk mitigation. Monitoring, evaluation, user education, and awareness are essential components. By fostering discussions and establishing guidelines, we can ensure the responsible and ethical development and use of LLMs and AI, maximizing benefits while minimizing risks.

## Background

The use of artificial intelligence (AI) tools, such as Large Language Models (LLMs), like ChatGPT (Chat Generative Pre-trained Transformer), Google Bard, and Bing AI, among others, is increasing in research publications [1]. LLMs are machine learning models specifically designed to handle Natural Language Processing (NLP) tasks, involving computer and human language interaction [2]. LLMs perform exceptionally in various NLP tasks, such as text generation, classification, summarization, question answering, and natural language inference. These models undergo training using extensive datasets comprising text and code, enabling them to grasp the intricate patterns of human language and generate coherent and grammatically accurate text; however, the response may not always be accurate [3]. For the sake of simplicity, we will refer to them in this article as “AI” or “LLMs.”

Authors (particularly non-native English speakers) can benefit enormously from AI to instantly deliver clear, compelling, and authentic writing. Furthermore, AI tools can potentially assist in several key areas, including writing, grammar check, language enhancement, reference citations, statistical analysis, formatting, and reporting standards compliance.  In addition, editors and publishers utilize AI-assisted tools for diverse purposes, including screening submissions for issues like plagiarism, image manipulation, and ethical concerns, triaging submissions, validating references, editing content, and adapting it for different publishing formats [4]. Additionally, these AI tools also play a crucial role in literature search and enhancing discoverability [5].

Initially, the enthusiastic voices acquired significant attention, but subsequently, the critical perspectives appear to be gaining increasing recognition [6, 7]. Therefore, in this article, our objective is to present a comprehensive overview of the application of AI in scientific writing. We will examine the benefits and drawbacks of this emerging technology, while also addressing the challenges faced by various stakeholders, including editors, publishers, academic institutions, and the wider research community.

**The Benefits and Drawbacks of AI in Research and Education**

The study of intelligent machines has a long history, and the debate over whether or not they can provide scientific insights has persisted since the advent of AI. Recently, there has been a renewed interest in the potential of AI in research and research policies. Like any novel tool, technique, or technology, AI brings about both positive and negative outcomes for research, researchers, and education involving both collective and individual utilization. AI is commonly regarded as beneficial in gathering information and providing support for impact and interdisciplinarity. However, utilizing AI as a means to expedite papers' production and metric-driven processes may amplify the negative aspects of academic culture.

The increasing involvement of AI in the writing, creation, and overall production of research papers has gained significant attention within the research and academic communities, and health organizations over the past year [8-10].

New AI technologies increased the risk of creating falsified work, combined with the difficulties of detecting such publications and the absence of perfect and comprehensive AI-detection technologies, increasing the risk of developing an environment encouraging fraudulent research. Risks associated with AI-generated research include the potential misuse of such work to manipulate and establish new healthcare policies, standards of care, and therapeutic interventions. Fabricating research using AI-based technology can be driven by various motives, including financial gain, desire for recognition, academic career advancement, and the need to enhance one's curriculum vitae, particularly among junior researchers facing intensifying competition [11]. While AI-based technologies can streamline routine research processes, they also risk contaminating the scientific research landscape and undermining the credibility of authentic works produced by other authors. Nevertheless, it is important to recognize that integrating AI in research should aid rather than replace human creativity [12]. **Figure 1** illustrates the proposed benefits and drawbacks of AI in research.

Moreover, the utilization of AI by students gives rise to several issues, posing challenges to fair student assessment, hindering student learning, and contributing to the spread of persuasive yet inaccurate essay tasks and homework assignments [13]. A study was conducted to investigate the potential impact of ChatGPT on academic dishonesty among students and reported that ChatGPT could be used to generate essays that meet specific requirements, potentially leading to students submitting plagiarized work [14]. The study also found that ChatGPT can be used to generate topics for essays, which could make it easier for students to find topics they are familiar with and can write about [14, 15]. The study concluded that ChatGPT poses a significant threat to the integrity of essay submissions in higher education environments [14, 15].

LLMs undergo training by utilizing extensive libraries of pre-existing texts. As a result, when provided with input from a human operator (e.g., prompt engineering), such as a question or seed text, LLMs generate responses or other outputs [16]. These outputs are essentially a combination of the training materials, adjusted based on the underlying algorithms. Due to their lack of consciousness [17], LLMs can only reproduce and reorganize existing information. Therefore, their statements can only be accidentally original, as they are unable to generate new thoughts [18].

ChatGPT and other LLM models offer numerous possibilities, such as condensing lengthy articles or generating initial versions of presentations that can be refined later [19]. In addition, they have the potential to assist researchers, students, and educators in brainstorming ideas [20].

Consequently, universities are required to adapt their teaching methods to the new and fast developments in AI [18]. With the rapid advancements in information technology and AI, there is a rising need for adaptable professionals, commonly known as T-shaped professionals, who have adequate knowledge of the utilities of AI in their specialty field [21]. The human is an enemy of what he is ignorant of; therefore, professionals and academicians should possess the capacity to go above their fear and disciplinary boundaries and exhibit a broad set of skills. In order to make optimal use of emerging technology, professionals and educators need to acquire proficiency in novel AI techniques and establish an advanced and efficient learning environment that smoothly incorporates AI.

Therefore, in order to effectively educate and advance AI literacy, teachers, academicians, researchers, and editors should ensure their proficiency in utilizing new technologies. Additionally, they should be knowledgeable about the limitations, drawbacks, and potential misuse of this technology while also understanding the advantages of the new development and how to ethically and appropriately leverage its advancements to enhance research and education.

**Discrimination between human and AI-generated text**

The growing prevalence of LLMs and anticipated advancements in their evolution will present a growing challenge in distinguishing between text authored by humans and that generated by machines. This inherent difficulty gives rise to a range of novel obstacles. For example, verifying human authorship without a doubt will become difficult, just as it will be challenging to expose machine-generated text falsely presented as original content definitively. Moreover, the utilization of LLM-generated text holds the potential for new forms of plagiarism, fraud, and the production of a large amount of false or misleading information [22].

Recently available data indicate that as LLMs continue to advance in sophistication, the task of differentiating between text produced by humans and that generated by AI becomes increasingly challenging. The growing complexity of LLMs obscures the distinctions, making it increasingly harder to determine the origin of the text as either human-generated or AI-generated, even by experts [23]. This poses significant challenges in accurately determining the authorship of the text and highlights the need for rigorous scrutiny and verification processes to ensure transparency and authenticity in written content [6].

This AI development carries important implications for upholding research integrity and underscores the necessity for enhanced methods and tools to identify instances of fraudulent research. The recent rise of paper mills and organizations driven by profit, and the production and selling of fraudulent manuscripts that imitate genuine research, has raised serious concerns and led to many retractions [24, 25]. The situation has become more complex with the advent of advanced writing and image-creation tools. AI can be used to create fraudulent papers, which is a problem because it undermines the integrity of science. While paper mills are intended to deceive, the use of AI may not always be intentional. However, the mere fact that AI can generate flawed ideas means that it is unscientific and unreliable, which should be a cause for concern for editors. Additionally, as Midjournary AI, DALL-E 2, Deep Dream Generator AI (OpenAI has developed a system that can generate realistic images and artwork based on textual descriptions provided to it in natural language) (**Figure 2**) [26], and other similar tools emerge, editors must establish journal policies regarding their usage and mandate the inclusion of content detection capabilities in these tools [27, 28].

These new developments have triggered ethical discussions on using AI in scholarly paper writing. While the unethical nature of generating fake papers is clear, there are legitimate concerns and debates about the implications of employing AI tools to aid in the creation of papers based on valid research data [29].

This raises several legitimate questions that need to be addressed before developing regulations for using AI in scientific and scholarly writings, among them; is it ethically acceptable for authors to utilize AI in writing scholarly articles?; is it possible and necessary to identify if AI has authored an article? If so, what is the significance?; what are the implications of AI-generated texts for plagiarism, including words, images, and ideas derived from algorithms trained on existing articles?; could peer review be exclusively conducted by dedicated AI tools capable of detecting fraud and validating data and figures?; and what considerations should be taken into account for equity and inclusion (i.e., the acceptable percentage), particularly for scholars with disabilities who may utilize AI tools as assistive or adaptive technology? On a related note, when a manuscript is published the conventional way, several people assist, including secretarial and technical support (e.g., lab technicians, graphic artists, administrative assistants, research assistance, and others) for conducting, overseeing, preparing, analyzing, typing, and formatting. Additionally, the use of built-in software for grammar checks, editing assistance from others (including peer reviewers and professional services), and so on. Hence, how far is AI-assisted delivery different from that of a conventional method of publication?

As AI-generated content becomes prevalent, there is a growing need for software that can detect plagiarism and identify scientific papers that have been written by AI in order to maintain the integrity of academic research. Plagiarism detection software was developed nearly two decades ago, but it took several years for most publishing companies to adopt it [30]. Generative AI-based software that produces text upon request is a new challenge, and there may not be enough time to implement effective measures to prevent the influx of computer-generated papers into the review process. Initial experiments using existing plagiarism detection software to identify AI-generated text have been unsuccessful and have shown that AI-generated essays might go undetected [15]. While these solutions may not eliminate the issue, they could potentially reduce the number of such papers entering the review process. New software called GPTZero, which was developed by a student at Princeton University, has received considerable attention in the media [31]. The software claims to be able to identify AI-generated text by analyzing two key features: perplexity and burstiness. Perplexity measures how similar the text is to what the AI language model has previously encountered, while burstiness measures the variability of sentences [32]. The author of GPTZero claims that the software has an exceptionally low false-positive rate, but its use remains controversial. GPTZero is not the first application designed to identify AI-generated writing; it is also unlikely to be the last. Many other applications have been developed too for this purpose, and new applications are being developed constantly. A study evaluated a method called DetectGPT for determining if a text passage was generated by a particular source model (GPT-3) [33]. DetectGPT was able to distinguish between human-written and LLM-generated text in over 95% of the cases when tested with several large language models. However, 95% is not 100%, and it is important to note that the number of potential LLMs is constantly increasing. Given that a text classifier capable of detecting text generated by AI is specific to a particular LLM model, the number of models that need to be tested may be a problem. However, the process of detecting AI-generated text is constantly evolving as AI-generation technology improves. To adequately respond to these advancements, detection systems must be able to analyze both similarities and differences in language across multiple levels, including syntax, semantics, and pragmatics [34]. Moreover, there are no studies on the accuracy of most used detection software, and major publishers or journals have not adopted it. Hence, it is ethically questionable to question the credibility of researchers solely based on an inadequately validated software detection algorithm. The main concern is that if a journal or publisher uses an un-validated AI-text detector and falsely accuses authors of using AI to generate their text, it could cause the author distress and harm. This is because a false positive outcome could damage the authors' reputation and make it difficult for them to publish their work in the future. Therefore, as long as there is a possibility that these text detectors could falsely accuse authors of cheating, they should be used with caution and with the understanding that their judgment could be wrong.

A final note on the software detection of similarities or AI written text needs to be addressed. Academic journal publishers and editors use similarity detection software to check if manuscripts are similar to other sources. They used to set acceptable similarity percentages, but they realized that human judgment is always needed. For example, a 3000-word manuscript that is 10% similar to another source may contain 300 words copied verbatim from that source, while a 30% similar manuscript may have scattered medical terminology and unintentional similarities [35]. Therefore, the detection software can identify the similarities but cannot determine the author's intention. Even if the software is developed to detect the use of AI-generated text, we cannot solely rely on the raw results [35]. Human judgment will still play a crucial role.

As a result, organizations like the Committee on Publication Ethics (COPE) [35], the World Association of Medical Editors (WAME) [10], and the Journal of American Medical Association (JAMA) Network journals [36], among others, have published recommendations to address the issues that editors, publishers, and the research community as a whole are confronting.

**AI authorship**

In January 2023, *Nature* reported on two preprints and two articles in the fields of science and health that listed ChatGPT as a named author [37]. Each of these publications included an affiliation for ChatGPT, and one of the articles even provided an email address for the nonhuman

"author" [37]. According to *Nature*, the inclusion of ChatGPT in the author byline of that particular article was acknowledged as an "error that will soon be corrected" [37]. However, these articles, and their nonhuman "authors" have already been indexed in databases like PubMed and Google Scholar.

Editors need to recognize that both Microsoft and OpenAI disclaim responsibility for the content created using their products. Microsoft does this with Word, and OpenAI does this with ChatGPT [18]. This means neither company is responsible for the accuracy, completeness, or legality of the content created using their products [38].

Then the question comes, are nonhuman AI, LLMs, machine learning, or similar technologies eligible for authorship? All currently available recommendations indicate that nonhuman AI and language models are not legible for authorship [18, 35, 36]. Therefore, the COPE statement indicated that AI tools are not eligible for authorship as they lack the ability to assume responsibility for the submitted work and approve copyright [35]. Moreover, being non-legal entities, they cannot confirm conflicts of interest or manage copyright and licensing agreements required by authors.

**The legitimate call for Authors' Declaration:**

Authors should disclose the use of artificial intelligence, LLMs, machine learning, or similar technologies in the acknowledgment section or, if applicable, in the methods section of formal research design or methods [36]. The disclosure should include a description of the created or edited content, as well as the name, version, and manufacturer of the language model or tool. On the other hand, publishers and journals should provide detailed instructions to authors about using AI for basic proofreading or editing; JAMA Networks journals guidelines excluded basic grammar, spelling, and reference-checking tools from the declaration [36].

According to the recommendations of WAME, it is advised that authors maintain transparency regarding the use of AI and LLMs and provide details on how they were utilized [18]. As the field is rapidly evolving, authors who employ AI in their paper-writing process should openly disclose this information and provide comprehensive technical specifications of the used LLMs, including its name, version, model, and source, as well as the method of application within the submitted paper, such as query structure and syntax [18]. This aligns with the ICMJE (International Committee of Medical Journal Editors) recommendation to acknowledge any writing assistance received [39].

Submitting and publishing content created by artificial intelligence, language models, machine learning, or similar technologies is discouraged by JAMA Network journals unless it is part of a formal research design or methods [36]. If such content is included, authors must clearly describe the generated content and provide the name, version, and manufacturer of the model or tool responsible for its creation. Finally, authors are responsible for ensuring the integrity of the generated content and its adherence to ethical standards.

**Concluding Remarks and Recommendations**

The substantial potential of AI in scientific writing is still not fully realized, urging scholars to give immediate attention to exploring its untapped avenues for research and discovery. However, to utilize its benefits responsibly, we must probe deeper into understanding the various utilities of AI applications while maintaining a delicate balance between ethical use, research excellence, and integrity, and avoiding misconduct.

Every application of AI comes with both advantages and disadvantages, necessitating a swift identification of key positive and negative aspects, especially concerning early career researchers. To ensure responsible use, it is imperative that all stakeholders, including developers, funders, universities, publishers, and researchers to engage in comprehensive discussions. Initiatives like the Declaration on Research Assessment (DORA) provide a foundation for promoting responsible metrics in research, and such efforts should be built upon [40].

Furthermore, addressing the demands of future research, anticipatory governance, and prediction is vital to encourage a research culture that adopts and regulates AI use in a beneficial and supportive manner. While data-related challenges are important, equal attention should be given to preserving the core principles and standards of research, such as curiosity, exploration, fascination, and integrity. Hence, determining the extent of AI integration in future research and its ultimate objectives becomes a critical decision that needs a global outlook.

As with any emerging technology, AI creates challenges for researchers, reviewers, editors, and the publishing community; however, its true benefits lie in how it is used. Stakeholder discussions are crucial for defining the beneficial and advantageous use of AI in research while considering situations where its adoption may be unsuitable or detrimental. This calls for international efforts to develop a consensus statement and detailed policies about the utility of AI in research.

Establishing guidelines and monitoring mechanisms that regulate AI as a supportive tool for genuine productivity, rather than a substitute for humans, is of the greatest importance. Further, developing scholarly publishing guidelines should involve input from diverse groups, including non-native English-speaking researchers who may use LLMS and AI to improve their writing style, as well as researchers with special needs who might need AI assistance, though this process may take time [41].

In the meantime, we propose the following recommendations for editors and authors; authors should openly disclose the use of artificial intelligence, LLMs, machine learning, or similar technologies in their research papers. This disclosure can be included in the acknowledgment section or, if applicable, in the methods section. A clear and transparent reporting of research methods and results enables readers to understand the study's conduct, evaluate the outcomes, prevent academic misconduct, and ensure the reproducibility of the research.

Furthermore, considering the aforementioned, numerous areas of limited knowledge exist, and potential avenues for further research regarding the involvement of AI in research are open. Likewise, there is a pressing demand for the development of practical and effective methods to identify and address instances of unethical AI utilization in research and publishing.

Finally, it is essential to regularly review and refine the developed editorial and publication policies to adapt to the evolving landscape of AI in scientific writing. By embracing responsible AI use and continuously improving guidelines, we can maximize the potential beneficial use of AI while upholding the principles of research and ensuring a fruitful future for scientific endeavors.

**Declaration:**

We hereby declare that the image included in Figure 2 was generated using the online version of Deep Dream Generator. Deep Dream Generator is a computer vision software developed by Google, utilizing a convolutional neural network to identify and enhance image patterns through algorithmic pareidolia.

**Legend of Figures**

**Figure 1**: The main benefits and drawbacks of AI in research.

**Figure 2:** This image was generated using an image-creation tool that produces realistic images and artwork from text descriptions. The original instructions requested "*an image featuring a friendly humanoid robot sitting at a messy desk with piles of scientific papers. The robot should be seen writing and printing scientific papers, with a computer displaying an edited paper, and additional elements like equations and charts on the papers to emphasize their scientific nature*." Through multiple regeneration attempts, we have arrived at this picture, which highlights that the current AI tools may not fully adhere to all the given instructions and text descriptions.

Figure 1:

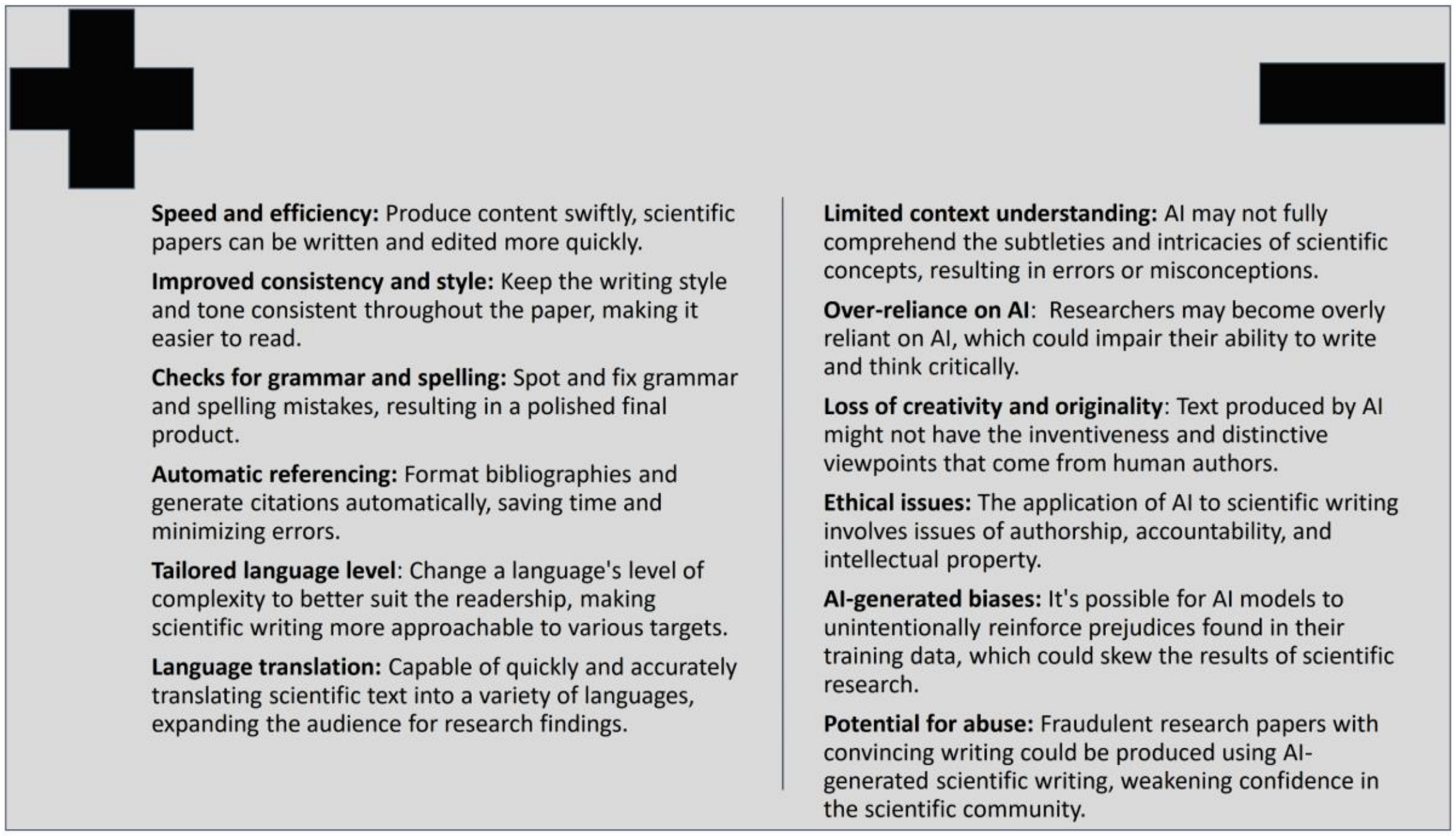
**Speed and efficiency:** Produce content swiftly, scientific papers can be written and edited more quickly.
**Improved consistency and style:** Keep the writing style and tone consistent throughout the paper, making it easier to read.
**Checks for grammar and spelling:** Spot and fix grammar and spelling mistakes, resulting in a polished final product.
**Automatic referencing:** Format bibliographies and generate citations automatically, saving time and minimizing errors.
**Tailored language level**: Change a language's level of complexity to better suit the readership, making scientific writing more approachable to various targets.
**Language translation:** Capable of quickly and accurately translating scientific text into a variety of languages, expanding the audience for research findings.
**Limited context understanding:** AI may not fully comprehend the subtleties and intricacies of scientific concepts, resulting in errors or misconceptions.
**Over-reliance on AI**: Researchers may become overly reliant on AI, which could impair their ability to write and think critically.
**Loss of creativity and originality**: Text produced by AI might not have the inventiveness and distinctive viewpoints that come from human authors.
**Ethical issues:** The application of AI to scientific writing involves issues of authorship, accountability, and intellectual property.
**AI-generated biases:** It's possible for AI models to unintentionally reinforce prejudices found in their training data, which could skew the results of scientific research.
**Potential for abuse:** Fraudulent research papers with convincing writing could be produced using AI-generated scientific writing, weakening confidence in the scientific community.

Figure 2:

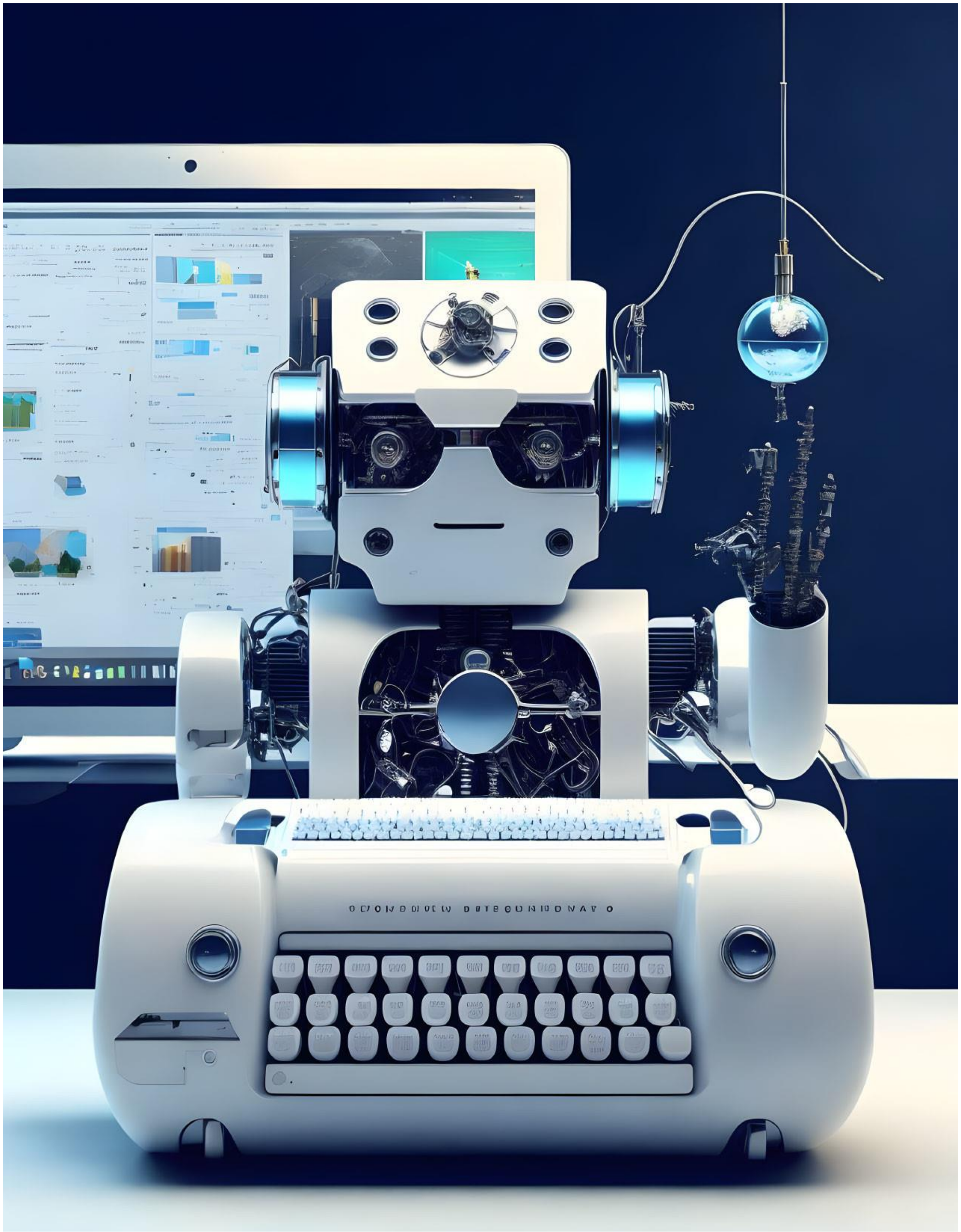